\documentclass[11pt,draftcls,onecolumn,journal]{IEEEtran}
\ifCLASSINFOpdf
\else
   \usepackage[dvips]{graphicx}
\fi
\usepackage{url}
\usepackage{cite}
\usepackage{xcolor}
\usepackage{cite}
\usepackage[linesnumbered,ruled,vlined]{algorithm2e}
\usepackage[font=small,labelfont=bf,labelsep=colon]{caption}
\usepackage{graphicx}   
\usepackage{subcaption} 

\usepackage{float} 
\usepackage{multicol}
\usepackage{amsmath}
\usepackage{amssymb}
\usepackage{algorithmic}
\usepackage{amsfonts}
\usepackage{amstext}
\usepackage{bigdelim}
\usepackage{bm}
\usepackage{boites,eurosym}
\usepackage{booktabs}
\usepackage{epsfig}
\usepackage{color}
\usepackage{epstopdf}
\usepackage{float}
\usepackage{graphicx}
\graphicspath{{PaperFigures/}}
\usepackage{booktabs}  
\usepackage{tabularx}  
\usepackage{multirow}  
\usepackage{siunitx}   
\usepackage{makecell}  
\usepackage{hyperref}
\usepackage{lipsum}
\usepackage{multirow}
\usepackage{pdfpages}
\usepackage{tabularx}
\usepackage{textcomp}
\usepackage{ulem}
\usepackage{xcolor}
\usepackage[numbers]{natbib}
\usepackage{cleveref}
\usepackage{cite}
\usepackage{siunitx}
\DeclareMathOperator*{\argmax}{arg\,max}

\newcommand{\ve}[1]{\boldsymbol{#1}}
\begin{document}

\title{Maximum likelihood for logistic regression model with incomplete and hybrid-type covariates
}

\author{M. Cherifi$^{*}$, X. Zhu$^{+}$, M. N. El Korso$^{+}$ and A. Mesloub$^{*}$%
\thanks{$^{*}$École Militaire Polytechnique, Algeria, $^{+}$  Paris Saclay University, France}
}

\maketitle

\begin{abstract}
Logistic regression is a fundamental and widely used statistical method for modeling binary outcomes based on covariates. However, the presence of missing data, particularly in settings involving hybrid covariates (a mix of discrete and continuous variables), poses significant challenges. In this paper, we propose a novel Expectation-Maximization based algorithm tailored for parameter estimation in logistic regression models with missing hybrid covariates. The proposed method is specifically designed to handle these complexities, delivering efficient parameter estimates. Through comprehensive simulations and real-world application, we demonstrate that our approach consistently outperforms traditional methods, achieving superior accuracy and reliability.
\end{abstract}

\begin{IEEEkeywords}
Maximum likelihood, logistic regression, parameter estimation, missing data.
\end{IEEEkeywords}

\section{Introduction}

In numerous practical applications, logistic regression serves as a fundamental method for modeling binary or categorical outcomes \citep{osborne2014best,leong2022logistic}. It is widely used in fields such as medicine, economics, and social sciences, where the goal is to predict the likelihood of an event or category based on observed features.  {However, real-world datasets frequently suffer from missing data  
leading to biased estimates of odds ratios and misclassification \cite{chen2022modified,raghunathan2004we,dung2021robust}, which in turn may compromise the validity of conclusions and decision-making processes in high‑stakes domains such as clinical risk prediction and policy analysis  
\cite{allison2009missing,Hippert2022}}. Nevertheless, the complexity escalates further when covariates are of mixed types (continuous and discrete), as traditional methods often fail to account for their interdependencies, leading to biased parameter estimates and reduced predictive accuracy \cite{raghunathan2004we, allison2009missing,audigier2016principal}. 

The analysis of mixed data with missing values has been the subject of extensive research. Among the pioneering approaches, the work of Olkin  \cite{olkin1961multivariate}, along with Krzanowski's extension \cite{little1985maximum}, laid the groundwork for maximum likelihood procedures. These methods rely on the general location model and utilize the Expectation-Maximization (EM) algorithm  \cite{dempster1977maximum,zhou2019} to estimate parameters in the presence of incomplete data. On the other hand, nonparametric methods have been proposed. A typical example is \textit{MissForest} \cite{stekhoven2012missforest}, which combines random forests with \textit{imputeFAMD} \cite{audigier2016principal}, an imputation method based on principal component analysis. However, these methods demonstrate reduced effectiveness when dealing with a large number of levels.  

While general-purpose imputation and maximum likelihood techniques have been widely studied for linear models, their direct application to logistic regression can be suboptimal due to the model’s non-linearity and the presence of mixed type data.  {As a result, specialized methods that explicitly account for the interplay between missingness and the logistic model structure are needed \cite{Martins2024,Cohen2023}.} 
Despite advances in handling missing data, the issue of missing covariates in logistic regression remains a significant challenge.  {Many existing methods predominantly focus on continuous covariates \cite{jiang2020logistic,Cherifi2025,Aladin2025,Lim2024,Alharthi2022}, limiting their applicability in practical scenarios where mixed types of covariates are present}. 
To overcome these limitations, we propose adopting the Stochastic Approximation Expectation-Maximization (SAEM) algorithm for parameter estimation in logistic regression with missing mixed-type covariates. The SAEM algorithm is well-suited for this task because it can handle complex missing data patterns and offers an adequate solution for handling missing covariates. By combining stochastic approximations with an iterative EM framework, SAEM provides a flexible and scalable approach for estimating parameters even when covariates are missing \cite{delyon1999convergence,zhou2020student}.
 In this paper, we develop and analyze a SAEM-based method for logistic regression that can handle missing data under ignorable mechanisms. Our approach is specifically tailored to deal with mixed-type covariates, leveraging the strengths of SAEM to provide adequate parameter estimates. Through extensive simulations and real-world applications, we demonstrate that the proposed method outperforms traditional techniques.

\section{Model setup}
\label{sec:Model}
In the following, we consider the logistic regression model for classification. Given a data set of $n$ points, for the $i$-th sample, we have the binary observation $y_i$ and the complete covariate values denoted by \(\bm{x}_i = [x_i^1, \ldots, x_i^l, x_i^{l+1}, \ldots, x_i^{h+l}]^T\). The first \(l\) variables of $\bm{x}_i$ are discrete, denoted by \(\bm{x}^d_i\), and each is defined on its associated space \(\mathbb{X}^d_j\) (for \(j = 1, \ldots, l\)). For simplicity, we enumerate the possible outcomes, allowing the outcome space to be represented by $\mathbb{X}^d_j = \{1,\ldots,M_j\}$, where $M_j$ corresponds to the number of levels. The remaining \(h\) components of the vector \(\bm{x}_i\) are continuous variables defined in \(\mathbb{R}^h\), denoted by \(\bm{x}^c_i\). The logistic model provides the statistical dependence between the binary class and the associated covariates:
{\begin{equation}
\log \left( \frac{p(y_i = 1 \mid \bm{x}_i)}{1 - p(y_i = 1 \mid \bm{x}_i)} \right) = \bm{\beta}^T \begin{bmatrix} 1  \ \bm{x}_i^T \end{bmatrix}^T,
\label{eq:log_odds}
\end{equation}
  in which \(\bm{\beta}\) represents the regression parameters.} 
 Due to missing values, some covariates are not always fully observed. Let $\bm{x}_{i,\text{obs}}$ represent the observed covariates and $\bm{x}_{i,\text{mis}}$ represent the missing covariates for the $i$-th sample. Furthermore, we define the collection of observed classes $\bm{y}=\{y_1,\ldots,y_n\}$ and denote the unobservable full dataset as $\mathcal{X} \triangleq \left\{\mathcal{X}_{\text{obs}}, \mathcal{X}_{\text{mis}}\right\}$ with $\mathcal{X}_{\text{obs}}= \left\{ \bm{x}_{i,\text{obs}}\right\}_i$ and
$\mathcal{X}_{\text{mis}}= \left\{ \bm{x}_{i,\text{mis}}\right\}_i$.

In this paper, we assume that the continuous data coordinates are sampled from a continuous distribution characterized by the probability distribution function {  \( \bm{x}^{c}_i \sim p(\bm{x}^{c}_i; \bm{\theta}^c) \)}, parametrized by the unknown vector \( \bm{\theta}^c \). Besides, discrete variables are sampled from a discrete distribution parametrized by $\bm{\theta}^d$. More precisely, each discrete variable $x_i^j$ is sampled from an independent discrete distribution, with the probability mass function 
{  \( x_i^j \sim p(x_i^j; \bm{\theta}^d_j) \)}.

\section{Algorithm design}
\subsection{Parameter inference}
Following the problem set up in the previous section, the unknown parameter of interest is grouped into \(\bm{\theta} = \left\{\bm{\theta}^c, \{\bm{\theta}^{d}_j\}_{j=1}^l, \bm{\beta}\right\}\). 
 For simplicity, we assume that continuous covariates follow a multivariate Gaussian distribution, while discrete covariates are modeled by independent multinomial distributions, which include Bernoulli distributions as special cases. As a result, we have \(\bm{\theta}^c = \{ \bm{\mu},\boldsymbol{\Sigma} \}\) for the continuous variables and \(\bm{\theta}^{d}_j = \{ \theta^{d}_j(m): m=1,\ldots,M_j \}\) for the $j$-th discrete variable, where $\theta^{d}_j(m)$ denotes the probability of level $m$. 
 Under the independence at the sample level and between discrete and continuous variables \cite{pruilh2024dynamic}, the complete likelihood is given by
{\small
\begin{multline}
\mathcal{L}(\bm{\theta}; \bm{y}, \mathcal{X}) = \sum_{i} L(\boldsymbol{\theta};y_i,\bm{x}_i)
= \sum_{i} \log\left(p(y_i,\bm{x}_i;\bm{\theta})\right)
\\ = \sum_{i} \left( y_i [1,\bm{x}_i^T] \bm{\beta} 
- \log\big(1 + \exp\big([1,\bm{x}_i^T] \bm{\beta}\big) +\sum_{j=1}^l \log \theta^{d}_j\left(x^j_i\right)\right.\\
\left. + \frac{1}{2} \big( -h \log(2 \pi) - \log |\bm{\Sigma}| 
- (\bm{x}_i^c - \bm{\mu})^T\bm{\Sigma}^{-1}(\bm{x}_i^c - \bm{\mu})\big) \right), 
\label{complet_loglik}
\end{multline}}
where $L(\boldsymbol{\theta};y_i,\bm{x}_i)$ is the log-likelihood of the $i$-th complete observation $(\bm{x}_i,y_i)$.
 Under the assumption of independence and ignorable missing data mechanism (meaning that the missing data does not depend on its value), the maximum log-likelihood estimator reads
\begin{equation}
\hat{\bm{\theta}} = \argmax_{\bm{\theta}} \sum_{i} \log p(y_i, \bm{x}_{i,\text{obs}}; \bm{\theta}),
\label{eq:log_likelihood}
\end{equation}
where the distribution $p(y_i, \bm{x}_{\text{i,obs}} ; \bm{\theta})$ of the observed data is given by marginalizing $p(y_i,\bm{x}_i;\bm{\theta})$ over the missing covariates. As the likelihood does not have a closed analytical form, it is impractical to maximize \eqref{eq:log_likelihood} directly. To this end, we propose applying the EM algorithm \cite{ibrahim1999missing}. 
{At the $t$-th iteration, given the current parameter estimates $\bm{\theta}^{(t)}$, the E-step of the reads refer to the computation of the expected complete log-likelihood function:
$Q^{(t)}(\bm{\theta}) = \mathbb{E}_{\mathcal{X}_{\text{mis}} \mid \bm{y}, \mathcal{X}_{\text{obs}}, \bm{\theta}^{(t-1)}} \left[ \mathcal{L}(\bm{\theta}; \bm{y}, \mathcal{X}) \right] = \sum_{i=1}^n Q_i^{(t)}(\bm{\theta}),$
where
\begin{align}
Q_i^{(t)}&(\boldsymbol{\theta}) 
=  \int  \sum_{\bm{x}_{i,\text{mis}}^d}
L(\boldsymbol{\theta}; y_i, \bm{x}_i) 
\, p(\bm{x}_{i,\text{mis}}^d \mid \bm{x}_{i,\text{mis}}^c,\bm{x}_{i,\text{obs}},y_i; \boldsymbol{\theta}^{(t-1)}) \notag \\
&\quad \times 
p(\bm{x}_{i,\text{mis}}^c \mid \bm{x}_{i,\text{obs}}, y_i; \boldsymbol{\theta}^{(t-1)}) \, \mathrm{d}\bm{x}_{i,\text{mis}}^c.\label{eq:EM}
\end{align} Then, the 
{M-step} maximizes the expected complete log-likelihood function with respect to the parameters to obtain updated estimates
$
\bm{\theta}^{(t)} = \arg\max_{\bm{\theta}} Q^{(t)}(\bm{\theta}).
$}

In the context of logistic regression, calculating the conditional expectation of the complete log-likelihood in \eqref{eq:EM} is challenging due to the intractability of the integral over the missing continuous variables. To address this, we apply the SAEM algorithm \cite{delyon1999convergence} to have an approximate stochastic solution. This consists of three steps: the S-step (simulation step), the SA-step (stochastic approximation update step), and finally, the M-step (maximization step). Specifically, for the \(t\)-th iteration:

\textbf{1) S-step} consists in sampling the missing continuous variables, i.e., for $i = 1, 2, \dots, n$, we sample:
\begin{equation}
\hat{\bm{x}}_{i,\text{mis}}^{c(t)} \sim p(\bm{x}_{i,\text{mis}}^c \mid \bm{x}_{i,\text{obs}}^{d}, \bm{x}_{i,\text{obs}}^c, y_i; \boldsymbol{\theta}^{(t-1)}).\label{target}
\end{equation}
To this end, we use the Metropolis--Hastings algorithm (please refer to 
the supplementary materials for more details). In the following, we denote $\hat{\bm{x}}^{c(t)}_{i}$ as the vector of $\bm{x}^{c}_{i}$ for which the missing value has been imputed using \eqref{target}. 

Given the imputed continuous variables and the statistical model presented in \Cref{sec:Model}, the posterior distribution of the missing discrete variables could be computed by

\begin{equation*}
\begin{split}
    &\tilde{w}_i^{(t)}(\bm{x}^d_{i,\text{mis}}) \triangleq p(\bm{x}_{i,\text{mis}}^d \mid \bm{x}_{i,\text{obs}}^d, \hat{\bm{x}}^{c(t)}_{i}, y_i; \boldsymbol{\theta}^{(t-1)})= \\
  &\frac{ p(y_i \mid \bm{x}_{i,\text{mis}}^d, \bm{x}_{i,\text{obs}}^d, \hat{\bm{x}}_{i }^{c(t)}; \boldsymbol{\beta}^{(t-1)}) \, p( \bm{x}_{i,\text{mis}}^{d}; \boldsymbol{\theta}^{d(t-1)}) 
 }
    {\sum_{\bm{x}_{i,\text{mis}}^d} 
    p(y_i \mid \bm{x}_{i,\text{mis}}^d, \bm{x}_{i,\text{obs}}^d, \hat{\bm{x}}_{i}^{c(t)}; \boldsymbol{\beta}^{(t-1)})p( \bm{x}_{i,\text{mis}}^{d}; \boldsymbol{\theta}^{d(t-1)})}, 
\end{split}
\end{equation*}
where $p(\bm{x}_{i,\text{mis}}^{d}; \boldsymbol{\theta}^{d(t-1)})$ is the distribution of the missing discrete covariates (a product of independent discrete distributions).

 \textbf{2) SA-step} updates the stochastic approximation $\hat{Q}^{(t)}$ of the objective function $Q^{(t)}$, that is,
{ \small
\begin{equation}\label{eq:Qupdate}
\hat{Q}^{(t)}(\boldsymbol{\theta}) = \hat{Q}^{(t-1)}(\boldsymbol{\theta}) + \delta_t \bigg(
  \sum_{i=1}^n \tilde{Q}_i^{(t)}(\bm{\theta}, \hat{\bm{x}}_{i,\text{mis}}^{c(t)}) 
    - \hat{Q}^{(t-1)}(\boldsymbol{\theta})
    \bigg),
\end{equation}}
{where \( \delta_{t} \) denotes a smoothing parameter (a decreasing sequence of positive values \cite{kuhn2004coupling}) and }
$
\tilde{Q}_i^{(t)}(\bm{\theta}, \hat{\bm{x}}_{i,\text{mis}}^{c(t)}) =
\sum_{\bm{x}_{i,\text{mis}}^d} L(\bm{\theta}; y_i, \bm{x}^d_{i}, \hat{\bm{x}}_{i}^{c(t)})\tilde{w}_i^{(t)}(\bm{x}^d_{i,\text{mis}}).$
Following \eqref{eq:EM}, $\tilde{Q}_i^{(t)}$ can be expressed as:

\begin{multline*}
\tilde{Q}_i^{(t)} ( \bm{\theta}, \hat{\bm{x}}_{i,\text{mis}}^{c(t)} ) 
    =   \\ 
    \sum_{\bm{x}_{i,\text{mis}}^d}  \tilde{w}_i^{(t)}(\bm{x}_{i,\text{mis}}^d)  
    \left( y_i [1,\hat{\bm{x}}_i^{(t)T}] \bm{\beta} - \log\left(1 + \exp\left([1,\hat{\bm{x}}_i^{(t)T}] \bm{\beta}\right)\right) \right) \\
    + \sum_{\bm{x}_{i,\text{mis}}^d} \tilde{w}_i^{(t)}(\bm{x}_{i,\text{mis}}^d) \log\left( p\left( \bm{x}_{i,\text{mis}}^{d} ; \bm{\theta}^{d} \right) \right)
    + \log \left( p\left( \bm{x}_{i,\text{obs}}^{d} ; \bm{\theta}^{d} \right) \right)\\
     -\frac{1}{2} \left(h \log(2 \pi) + \log |\bm{\Sigma}| 
    +(\hat{\bm{x}}_i^{c(t)} - \bm{\mu})^T\bm{\Sigma}^{-1}(\hat{\bm{x}}_i^{c(t)} - \bm{\mu})\right),      
\end{multline*}
 where \(\hat{\bm{x}}_i^{(t)} = \left[(\bm{x}_{i,\text{mis}}^d)^T,(\bm{x}_{i,\text{obs}}^d)^T,(\hat{\bm{x}}^{c(t)}_i)^T \right]^T \).

We observe that $\tilde{Q}^{(t)}_i$ is composed of three separate parts, each of which involves only the parameters $\bm{\beta}$, $(\bm{\mu},\bm{\Sigma})$, and $\bm{\theta}^d$, respectively. Hence, the update of $\hat{Q}^{(t)}$ could be performed component-wisely. More precisely, we define the sum across the number of samples for each component of $\tilde{Q}^{(t)}_i$ as
{\scriptsize 
\begin{equation*}
\begin{split}
    &\tilde{G}^{(t)}_1(\bm{\beta}) = \sum_{i=1}^{n}
    \sum_{\bm{x}_{i,\text{mis}}^d}
    w_{i,\text{mis}}^{(t-1)}(\bm{x}_{i,\text{mis}}^d) \cdot 
    \log \left( p\left(y_i \mid
    \bm{x}_{i}^d, \hat{\bm{x}}^{c(t)}_i; \boldsymbol{\beta} \right) \right),\\ &\tilde{G}^{(t)}_2(\bm{\theta}^d)=\sum_{i=1}^{n}\sum_{\bm{x}_{i,\text{mis}}^d} \tilde{w}_{i}^{(t)}(\bm{x}_{i,\text{mis}}^d) \log\left( p\left( \bm{x}_{i,\text{mis}}^{d} ; \bm{\theta}^{d} \right) \right) 
    +
    \log \left( p\left( \bm{x}_{i,\text{obs}}^{d} ; \bm{\theta}^{d} \right) \right),\\
    &\tilde{G}^{(t)}_3(\bm{\mu},\bm{\Sigma}) = -\frac{n}{2}\log|\Sigma| - \frac{1}{2}\operatorname{Tr} (\tilde{T}_{2}^{(t)}\Sigma^{-1} ) -\frac{1}{2}\bm{\mu}^T\Sigma^{-1}(\bm{\mu} - 2 \tilde{T}^{(t)}_{1}), 
    \end{split}
\end{equation*}}
where we ignore the constant term and use the sufficient statistics $\tilde{T}_{1} = \frac{1}{n} \sum_{i=1}^n \hat{\bm{x}}^{c(t)}_i$ and $\tilde{T}_{2}=\sum_{i=1}^n \hat{\bm{x}}^{c(t)}_i (\hat{\bm{x}}^{c(t)}_i)^T$ of multivariate Gaussian distribution to simplify the expression of $\tilde{G}^{(t)}_3$. 
By adapting the same updating procedure in \eqref{eq:Qupdate} induced by $\tilde{G}_k$, we define
\begin{equation}
    G^{(t)}_k(\cdot) = G^{(t-1)}_k(\cdot) + \delta_t(\tilde{G}^{(t)}_k(\cdot) - G^{(t-1)}_k(\cdot)), \,k=1,2,3,
\end{equation}
with $G^{(1)}_k=\tilde{G}^{(1)}_k$. Therefore, $\hat{Q}^{(t)}$ could be decomposed as
\begin{equation}\label{eq:Qdecomp}
    \hat{Q}^{(t)}(\bm{\theta}) = G^{(t)}_1(\bm{\beta}) + G^{(t)}_2(\bm{\mu},\bm{\Sigma}) + G^{(t)}_3(\bm{\theta}^d).
\end{equation}
Importantly, because $\tilde{G}^{(t)}_3$ is linear in $\tilde{T}_{1}$ and $\tilde{T}_{2}$, $G^{(t)}_3$ could be expressed in the same form
{\small
\begin{equation*}
    G_3(\bm{\mu},\bm{\Sigma}) = -\frac{n}{2}\log|\Sigma| - \frac{1}{2}\operatorname{Tr} (T_{2}^{(t)}\Sigma^{-1} ) -\frac{1}{2}\bm{\mu}^T\Sigma^{-1}(\bm{\mu} - 2 T^{(t)}_{1}),
\end{equation*}}
with updated sufficient statistics
\begin{align*}
    T_{1}^{(t)} &= T_{1}^{(t-1)} + \delta_t \left(\Tilde{T}^{(t)}_{1} - T_{1}^{(t-1)}\right), \\
    T_{2}^{(t)} &= T_{2}^{(t-1)} + \delta_t \left(\Tilde{T}^{(t)}_{2} - T_{2}^{(t-1)}\right). 
\end{align*}
Similarly, $G^{(t)}_2$ takes the same form as $\tilde{G}^{(t)}_2$ 
with the posterior distribution $\tilde{w}_{i}^{(t)}$ replaced by the updated distribution $w_{i}^{(t)}$ given by
\begin{equation*}
    w_{i}^{(t)}(\cdot) = w_{i}^{(t-1)}(\cdot) + \delta_t\left(\tilde{w}_{i}^{(t)}(\cdot)-w_{i}^{(t-1)}(\cdot)\right).
\end{equation*}

 
 \textbf{3) M-step} updates the estimation of \(\boldsymbol{\theta}\) as
$\boldsymbol{\theta}^{{(t)}} = \argmax_{\boldsymbol{\theta}} \hat{Q}^{(t)}(\boldsymbol{\theta})$. According to \eqref{eq:Qdecomp}, the maximization could be conducted separately. 
For the regressor parameters $\bm{\beta}$, we use a gradient descent-based method to solve
\begin{equation}\label{M_regressor_parameters}
    \hat{\boldsymbol{\beta}}^{(t)} = \argmax_{\boldsymbol{\beta}}  G_1^{(t)}({\bm{\beta}}). 
\end{equation}	

The parameter update for the continuous distribution could be computed analytically by exploring the sufficient statistics
\begin{equation}
\hat{\bm{\mu}}^{(t)} = T_{1}^{(t)}, \quad 
\hat{\bm{\Sigma}}^{(t)} = \frac{T_{2}^{(t)}}{n} - T_{1}^{(t)} \left(T_{1}^{(t)}\right)^T.
\end{equation}
    

Finally, the parameters of the discrete variables are estimated as
\begin{equation}
    \bm{\theta}^{d(t)} = \argmax_{\bm{\theta}^d \in \bm{\Theta}^d} G^{(t)}_2(\bm{\theta}^d),
\end{equation}
where $\bm{\Theta}^d$ is the space of valid probability mass function of the discrete variables involved in the analysis. By exploring the independent structure of the discrete variables, the estimation has the following analytical form \newline
$    \theta^{d(t)}_j(m) =  n^{-1} ( |\{i: x^j_{i,\text{obs}}=m\}| + \sum_{i \in I_{j,\text{mis}}}w^{j(t)}_i(m)), $
where $I_{j,\text{mis}}$ denotes the set of indices corresponding to samples with a missing $j$-th covariate, and $w^{j(t)}_i$ represents the marginalized form of $w^{(t)}_i$ for the $j$-th component.

\subsection{Prediction in the Presence of Missing Data} \label{pred_MCAR}
This section presents a natural procedure to make predictions with potentially missing values in mixed data during the testing phase, assuming that training and test data come from the same distribution. 
The class prediction reads

\begin{equation*}
\hat{y}_i   =   \argmax_{y_i\in \{0,1\}} p(y_i \mid \bm{x}_{i,\text{obs}}; \hat{\bm{\theta}}).
\end{equation*}
where $p(y_i \mid \bm{x}_{i,\text{obs}}; \hat{\bm{\theta}})$ reads as the following marginalization 
\begin{align}
    p(y_i \mid \bm{x}_{i,\text{obs}}; \hat{\bm{\theta}}) & =\int_{\bm{x}^{c}_{i,\text{mis}}} \sum_{\bm{x}_{i,\text{mis}}^d} p\left(y_i, \bm{x}^{c}_{i,\text{mis}}, \bm{x}_{i,\text{mis}}^d \mid \bm{x}_{i,\text{obs}} ; \hat{\bm{\theta}}\right) \, d\bm{x}_{i,\text{mis}}^c \\
    & =\mathbb{E}_{\bm{x}_{i,\text{mis}}\mid \bm{x}_{i,\text{obs}};\hat{\bm{\theta}}}\left[p(y_i\mid \bm{x}_{i,\text{mis}},\bm{x}_{i,\text{obs}};\hat{\bm{\beta}})\right].\label{eq:marginalization}
\end{align}

By leveraging the structure of the statistical model, we could factorize the conditional distribution of $\bm{x}_{i,\text{mis}}$ by
\begin{equation*}
p(\bm{x}_{i,\text{mis}}\mid\bm{x}_{i,\text{obs}};\hat{\bm{\theta}}) = p(\bm{x}^d_{i,\text{mis}};\hat{\bm{\theta}}^d)p(\bm{x}^c_{i,\text{mis}}|\bm{x}^c_{i,\text{obs}};\hat{\bm{\theta}}^c).
\end{equation*}
As a result, the expectation in \eqref{eq:marginalization} could be evaluated separately for the missing discrete and continuous covariates. For discrete variables, which have a finite number of possible values, the associated expectation is computed as a weighted sum. For continuous variables, we approximate this expectation using Monte Carlo simulations, generating $S$ samples $\left\{\bm{x}^{c(s)}_{i,\text{mis}}\right\}_{s=1}^S$from the conditional distribution $p(\bm{x}^c_{i,\text{mis}}|\bm{x}^c_{i,\text{obs}};\hat{\bm{\theta}}^c) $. In summary, the final prediction reads

{\small
\begin{equation*}
\tilde{y}_i =   \argmax_{y_i \in \{0,1\}} \sum_{\bm{x}_{i,\text{mis}}^d} p\big(\bm{x}_{i,\text{mis}}^d ; \hat{\bm{\theta}}^d\big) \sum_{s=1}^S
p\big(y_i \mid \bm{x}^{c(s)}_{i,\text{mis}}, \bm{x}_{i,\text{mis}}^d, \bm{x}_{i,\text{obs}} ; \hat{\bm{\beta}}\big).
\end{equation*}}

\section{Numerical examples}
\textit{Synthetic Dataset: } We generate a synthetic dataset composed of \(n = 1000\) samples with \(7\) covariates. Each sample includes a binary response variable \(y_i \in \{0,1\}\) and a set of covariates \(\boldsymbol{x}_i\), consisting of both discrete and continuous variables. The first covariate, \(x_i^1\), is binary and generated from a Bernoulli distribution with a success probability of \(0.5\), representing two possible outcomes. The second covariate \( x_i^2 \) is multinomial, taking one of the five possible values in the set \( \{1, 2, 3, 4, 5\} \), sampled from a multinomial distribution with probabilities \([0.1, 0.2, 0.3, 0.25, 0.15]\). The remaining covariates, \(x^3_i, \dots, x^7_i\), are continuous and simulated from a multivariate normal distribution with a mean vector \(\bm{\mu} = \bm{0}\) and a covariance matrix \newline
$ {\color{white}{p}}\ \ \ \ \ \  \ \ \ \ \ \ \hspace{0.25cm}
\bm{\Sigma} = \begin{pmatrix}
4 & 0.5 & 0.2 & 0.1 & 0.3 \\
0.5 & 3 & 0.1 & 0.3 & 0.2 \\
0.2 & 0.1 & 2 & 0.4 & 0.5 \\
0.1 & 0.3 & 0.4 & 3 & 0.2 \\
0.3 & 0.2 & 0.5 & 0.2 & 5
\end{pmatrix},
$ \newline
representing the dependencies and variations within the continuous covariates.
To simulate the binary response variable \( y_i \), we use the coefficient vector 
$ \bm{\beta} = [0, -0.9, 0.01, 0.1, -0.6, 0.3, 0.01, 0.8]^{T} $ in a logistic regression model. The probability of success for each sample is computed using the logistic function; 
 Based on this probability, the values of \( y_i \) are drawn from a
Bernoulli distribution. To simulate missing data, we randomly introduced missingness in the covariates 
 under different missing data mechanisms. 
{For the smoothing parameter $\delta_t$, we use the step sequence \(
\delta_t = (t + t_1)^{-\tau}
\) with $t_1 = 50$. During the first 50 iterations, we set $\delta_t = 1$ to favor the MLE, and thereafter $\delta_t$ decreases with exponent $\tau = 1$, providing a balance between speed and stability.}
\begin{figure}[ht]
    \centering
    \includegraphics[width=0.55\textwidth,height=0.30\textwidth,keepaspectratio]{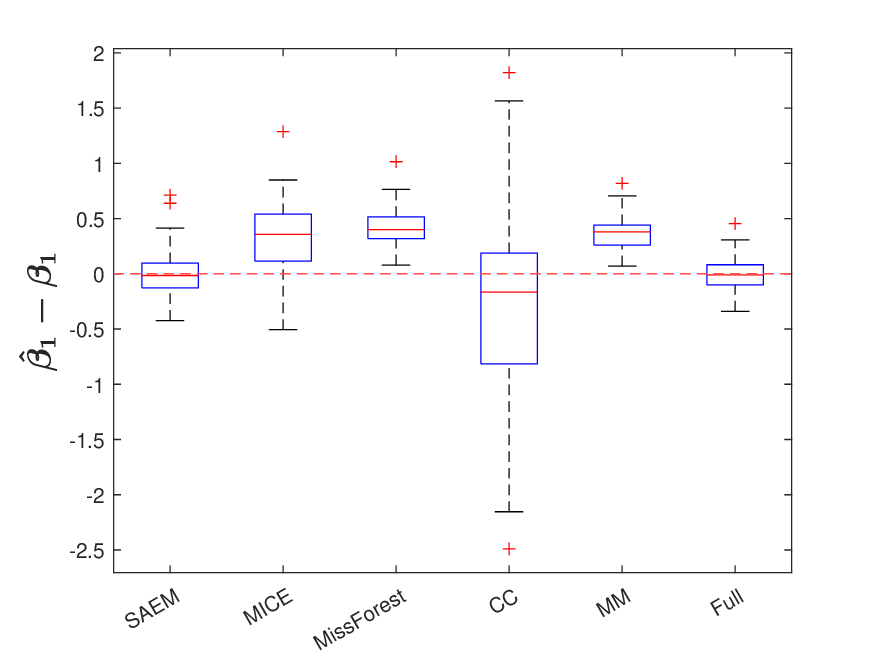} 
    \caption{{Boxplot of the difference between the estimate and the ground truth for \( \beta_{1} \) over  {100 algorithm runs} with 30\% MAR missing data mechanisms.}}
    \label{fig:bias_mar}
\end{figure}
{The proposed method is compared with the \textit{MICE} approach (Multiple Imputation by Chained Equations), adapted to handle mixed data \cite{van2011mice}, and the \textit{Miss-Forest} algorithm, which has also been tailored for mixed data \cite{stekhoven2012missforest}. Additionally, we include the \textit{MeanMode (MM)} method \cite{zhang2016missing}, which imputes missing values for continuous variables using the mean of the observed values and for discrete variables using their mode. We also consider the \textit{Complete Case (CC)} approach, which removes samples containing at least one missing value. For all these methods, the training and test datasets are processed separately before being passed to the logistic regression model. Finally, the considered methods include the analysis of cases without missing data (\textit{Full}) serving as a reference.
To simulate a Missing at Random (MAR) mechanism, we modeled missingness probabilities for $x_1$, $x_2$, $x_4$, and $x_6$ via logistic regression using only the fully observed variables $x_3$, $x_5$, and $x_7$. Regression coefficients were tuned to control the target missingness rate. Probabilities were then converted to binary indicators through Bernoulli draws, replacing values with \texttt{NA} where indicated.}
{
First, we focus on the accuracy of estimating \( \bm{\beta} \). To this end, we study the biases of the estimates for each component.  
Fig.~\ref{fig:bias_mar} illustrates the difference between the estimate and the true value for the first component of the parameter vector of interest, \( \beta_1 \) ( {in which algorithm runs, means simulations,   each with random spiting of training and testing samples and random missingness generation in the case of synthetic missingness}). In all cases, the proposed algorithm demonstrates a clear advantage, consistently achieving lower bias compared to the \textit{MICE} and \textit{MissForest} algorithms. Furthermore, the \textit{MM} and \textit{CC} methods exhibit even higher bias, highlighting their limitations compared to more advanced approaches.  {
Additional results on the RMSE of~\(\bm\beta\) and on predictive performance are given in the supplementary materials (Section~\ref{sec:Aresults}); note that we omitted the \textit{CC} method from the prediction boxplots—since, as shown in Fig~\ref{fig:bias_mar} and ~\ref{fig:norm}, \textit{CC} exhibits a catastrophic bias that would compress the plot scales and mask the differences between methods. The source code for these simulations is available in \cite{cherifi_mixed_2025}.}} 
{Regarding the time complexity of the proposed algorithm, we report the prediction time   in the Supplementary Materials. We can notice that the proposed algorithm is faster
than the MissForest by a significantly larger margin (e.g., almost 30 faster for 200
samples as shown in the Supplementary Materials).}



{\textit{Real Dataset:} In this section, we assess the predictive performance of the \textit{SAEM} algorithm on four real‐world datasets featuring both continuous and categorical predictors and a binary response. We employ the \textit{Heart Disease} dataset from the UCI Machine Learning Repository~\cite{heart_disease_45}, the \textit{Algerian Forest Fire} dataset also from the UCI Machine Learning Repository~\cite{algerian_forest_fires_547}, the \textit{Titanic} dataset from Kaggle~\cite{titanic_dataset}, and the \textit{Mtcars} dataset originally compiled by Henderson and Velleman~\cite{Henderson1974}. Detailed descriptions of these datasets are provided in the supplementary materials. The \textit{Heart Disease} and \textit{Titanic} datasets contain naturally occurring missing values at rates of 19\% and 12\%, respectively, whereas for the \textit{Algerian Forest Fire} and \textit{Mtcars} datasets we have artificially introduced missing values under a MCAR mechanism.
Tables~\ref{tab:Heart Disease} and~\ref{tab:metrics_comparison_20_40_ALgerian} present the performance comparison of various methods on the \textit{Heart Disease} dataset with 19\% naturally missing values and the \textit{Algerian Forest Fires} dataset with artificially injected MCAR missingness at rate of 
40\%, respectively. As shown in Table~\ref{tab:Heart Disease}, our proposed \textit{SAEM} approach consistently outperforms the competing methods, achieving superior predictive accuracy in the presence of naturally occurring missing data. Similarly, in Table~\ref{tab:metrics_comparison_20_40_ALgerian}, \textit{SAEM} demonstrates robust performance across both missingness levels, maintaining its advantage over alternative methods, including \textit{MICE} and \textit{MissForest}. The \textit{MICE} method performs poorly on both datasets, especially at higher levels of missingness, due to its underlying assumption of linearity, which hinders its ability to capture complex, non-linear relationships present in real-world data. Although the \textit{MissForest} method yields acceptable results, it remains inferior to \textit{SAEM}, which consistently delivers the best performance. These results underscore the effectiveness and robustness of our \textit{SAEM} approach in handling diverse missing data scenarios, whether naturally occurring or artificially introduced.
Results pertaining to the \textit{Titanic} and \textit{Mtcars} datasets are presented in the supplementary materials (Section~\ref{sec:Aresults}).
}

\begin{table}[h!]
\centering
\scriptsize 
\begin{tabular}{lcccc}
\toprule
Metric        & SAEM                 & MICE             & MM                   & MissForest       \\
\midrule
AUC           & \textbf{0.859} (0.041)        & 0.839 (0.060)    & \textbf{0.858 (0.036)} & 0.856 (0.040)    \\
Accuracy      & \textbf{0.833 (0.035)} & 0.795 (0.061)    & 0.770 (0.094)        & 0.818 (0.040)    \\
Precision     & \textbf{0.795 (0.073)} & 0.730 (0.105)    & 0.779 (0.164)        & 0.780 (0.082)    \\
Sensitivity   & \textbf{0.724 (0.067)} & 0.711 (0.104)    & 0.613 (0.234)        & 0.698 (0.099)    \\
Specificity   & \textbf{0.895 (0.039)} & 0.842 (0.104)    & 0.862 (0.180)        & 0.887 (0.052)    \\
F1 score      & \textbf{0.755 (0.054)} & 0.712 (0.075)    & 0.636 (0.166)        & 0.730 (0.067)    \\
Brier score   & \textbf{0.132 (0.022)} & 0.155 (0.040)    & 0.173 (0.061)        & 0.143 (0.029)    \\
\bottomrule
\end{tabular}
\caption{{Performance comparison of methods on the Heart Disease dataset with 19\,\% naturally missing data (mean ± standard deviation) over  {100 algorithm runs}.}}
\label{tab:Heart Disease}
\end{table}

\begin{table}[h!]
\centering

   \begin{tabular}{lcccc}
Metric & SAEM (40\%) & MICE (40\%) & MM (40\%) & MissForest (40\%) \\ \hline
AUC & \textbf{0.970} (0.022) & 0.787 (0.138) & 0.962 \textbf{(0.021)} & 0.950 (0.034) \\
Accuracy & \textbf{0.930 (0.029)} & 0.741 (0.123) & 0.892 (0.038) & 0.919 (0.034) \\
Precision & \textbf{0.933 (0.042)} & 0.766 (0.115) & 0.911 (0.053) & 0.922 (0.043) \\
Sensitivity & \textbf{0.942 (0.035)} & 0.767 (0.148) & 0.896 (0.077) & 0.934 (0.049) \\
F1 score & \textbf{0.937 (0.028)} & 0.762 (0.120) & 0.900 (0.041) & 0.927 (0.033) \\
Brier score & \textbf{0.057 (0.023)} & 0.247 (0.126) & 0.090 (0.035) & 0.077 (0.035) \\ \hline
\end{tabular}
\caption{Performance comparison of methods on the \textit{Algerian Forest Fires} dataset with 
40\% MCAR missing data (mean ± standard deviation) over  {100 algorithm runs}.}
\label{tab:metrics_comparison_20_40_ALgerian}
\end{table}

\section{Conclusion}

In this paper, we proposed a novel approach based on the {SAEM} algorithm for parameter estimation in logistic regression models with missing mixed-type covariates. 
  The results of our simulations and real data applications demonstrate that the SAEM approach outperforms classical techniques, offering more robust and reliable parameter estimates, even in challenging missing data situations. 

\newpage

\newpage
\begin{center}
    \textbf{Supplementary Materials} 
\end{center}
\renewcommand{\thesection}{S.\Roman{section}} 
\setcounter{section}{0}

\section{Metropolis--Hastings algorithm}
\label{sec:MHalgo}
In logistic regression for mixed data, it is generally not feasible to sample the unobserved continuous data directly from the conditional distribution \eqref{target}  because it requires evaluating an integral that cannot be solved analytically. One approach to address this is to employ the Metropolis--Hastings (MH) algorithm, which generates a Markov chain with the stationary distribution matching the target distribution. After \(S\) iterations (i.e., burn-in), the values of this Markov chain are considered as approximate samples from the target distribution.
 
The conditional distribution of the continuous missing covariates is given by
\begin{equation}
   \label{eq:condiXc}
p(\bm{x}^{c}_{i,\text{mis}} \mid y_i, \bm{x}_{i,\text{obs}};\boldsymbol{\theta}) = \frac{
\sum_{\bm{x}_{i,\text{mis}}^d}  
p( y_i, \bm{x}_{i,\text{mis}}^d, \bm{x}^{c}_{i,\text{mis}}|\bm{x}_{i,\text{obs}} ; \boldsymbol{\theta})
}{ 
p( y_i \mid \bm{x}_{i,\text{obs}} ; \boldsymbol{\theta})
}. 
\end{equation} 
From the setup of the statistical model in \Cref{sec:Model}, each element in the denominator could be factorized as

\begin{equation}
\label{eq:condiXcFact}
p( y_i, \bm{x}_{i,\text{mis}}^d, \bm{x}^{c}_{i,\text{mis}}|\bm{x}_{i,\text{obs}} ; \boldsymbol{\theta}) = p(y_i|\bm{x}_{i,\text{mis}}^d, \bm{x}^{c}_{i,\text{mis}},\bm{x}_{i,\text{obs}};\ve{\beta}) \\
\times p(\bm{x}^d_{i,\text{mis}};\ve{\theta}^d)p(\bm{x}^c_{i,\text{mis}}|\bm{x}^c_{i,\text{obs}};\ve{\theta}^c).
\end{equation} 

Combining \eqref{eq:condiXc} and \eqref{eq:condiXcFact} yields
{\small
\begin{equation*}\label{eq:posterior}
\begin{split}
   & p(\bm{x}^{c}_{i,\text{mis}} \mid y_i, \bm{x}_{i,\text{obs}};\boldsymbol{\theta}) \propto f(\bm{x}^{c}_{i,\text{mis}};\ve{\theta}) = \\
&\sum_{\bm{x}_{i,\text{mis}}^d}  p(y_i \mid \bm{x}^{c}_{i,\text{mis}}, \bm{x}_{i,\text{mis}}^d, \bm{x}_{i,\text{obs}}; \boldsymbol{\beta}) p(\bm{x}^d_{i,\text{mis}};\ve{\theta}^d) p(\bm{x}^{c}_{i,\text{mis}} \mid \bm{x}^{c}_{i,\text{obs}}; \ve{\theta}^c).
\end{split}
\end{equation*}}

Given the explicit form of $f(\bm{x}^{c}_{i,\text{mis}};\ve{\theta})$, we could apply the Metropolis--Hastings algorithm to sample $p(\bm{x}^{c}_{i,\text{mis}} \mid y_i, \bm{x}_{i,\text{obs}};\boldsymbol{\theta})$, as summarized in \Cref{alg:MH}. The proposal distribution is set to \( g(\bm{x}_{i,\text{mis}}^c ;\bm{\theta}^c ) = p(\bm{x}_{i,\text{mis}}^c \mid \bm{x}_{i,\text{obs}}^c; \bm{\mu}, \bm{\Sigma}) \), which is a multivariate normal distribution with the mean and variance \(\bm{\mu}_i, \bm{\Sigma}_i \) given by
\begin{equation}
\bm{\mu}_i = \bm{\mu}_{i,\text{mis}} + \bm{\Sigma}_{i,\text{mis},\text{obs}} \bm{\Sigma}^{-1}_{i,\text{obs},\text{obs}}(\bm{x}_{i,\text{obs}} - \bm{\mu}_{i,\text{obs}}),
\end{equation}
\begin{equation}
\bm{\Sigma}_i = \bm{\Sigma}_{i,\text{mis},\text{mis}} - \bm{\Sigma}_{i,\text{mis},\text{obs}} \bm{\Sigma}^{-1}_{i,\text{obs},\text{obs}} \bm{\Sigma}_{i,\text{obs},\text{mis}}.
\end{equation}

\begin{algorithm}[ht]
\small
\caption{Metropolis–Hastings Sampling}
\label{alg:MH}
\KwIn{Initial sample $\bm{\theta}_0$, proposal distribution $g(\cdot)$, target distribution $f(\cdot)$, number of iterations $S$}
\KwOut{Samples $\{\bm{x}_{i,\mathrm{mis}}^{(s)}\}_{1 \le i \le n,\ 1 \le s \le S}$}
\BlankLine
Initialize $\bm{x}_{i,\mathrm{mis}}^{(0)} \sim g(\cdot)$\;
\For{$s \leftarrow 1$ \KwTo $S$}{
  Propose $\bm{x}_{i,\mathrm{mis}}^{(s)} \sim g(\cdot)$ and draw $u \sim \mathcal{U}(0,1)$\;
  Compute
  \[
    r \;=\;
    \frac{f\bigl(\bm{x}_{i,\mathrm{mis}}^{(s)}\bigr)\,g\bigl(\bm{x}_{i,\mathrm{mis}}^{(s-1)}\bigr)}
         {f\bigl(\bm{x}_{i,\mathrm{mis}}^{(s-1)}\bigr)\,g\bigl(\bm{x}_{i,\mathrm{mis}}^{(s)}\bigr)}.
  \]
  \eIf{$u < r$}{
    Accept (keep $\bm{x}_{i,\mathrm{mis}}^{(s)}$)\;
  }{
    Reject: $\bm{x}_{i,\mathrm{mis}}^{(s)} \leftarrow \bm{x}_{i,\mathrm{mis}}^{(s-1)}$\;
  }
}
\end{algorithm}

\begin{table}[h!]
\centering
\caption{Mean Bias and RMSE of \(\bm{\beta}\) over  {100 algorithm runs} with 50\% MCAR. Best values among SAEM, MICE, and MissForest are highlighted in \textbf{bold}.}
\scalebox{0.86}{%
\begin{tabular}{cccccc}
\toprule
\(\beta\) & Metric & SAEM & MICE & MissForest & Full \\
\midrule
\(\beta_1\) & Bias       & \textbf{-0.0143}  & -0.4118  & -0.1376  & 0.0374  \\
           & RMSE       & 0.4397           & 3.6966           & \textbf{0.3323}                   & 0.2618 \\
\midrule
\(\beta_2\) & Bias       & -0.0847  & 0.0949   & \textbf{-0.0180}  & 0.0045  \\
           & RMSE       & \textbf{0.3446}                   & 2.5319           & 0.5699          & 0.1752 \\
\midrule
\(\beta_3\) & Bias       & \textbf{-0.0006} & 0.1635  & -0.0077 & -0.0079 \\
           & RMSE       & 0.1354          & 1.1545           & \textbf{0.0906}                   & 0.0745 \\
\midrule
\(\beta_4\) & Bias       & \textbf{0.0130}  & 0.1386  & -0.0339  & 0.0025  \\
           & RMSE       & 0.0860          & 0.63372           & \textbf{0.0689}                   & 0.0437 \\
\midrule
\(\beta_5\) & Bias       & \textbf{-0.0337} & -0.0826 & 0.2462   & -0.0095 \\
           & RMSE       & \textbf{0.1059}                   & 0.7330  & 0.2567                   & 0.0627 \\
\midrule
\(\beta_6\) & Bias       & \textbf{0.0399}  & -0.3037 & -0.1598  & 0.0092  \\
           & RMSE       & \textbf{0.1354}          & 1.0022           & 0.1811                   & 0.0571 \\
\midrule
\(\beta_7\) & Bias       & \textbf{0.0009}  & 0.1248  & 0.0079   & 0.0067  \\
           & RMSE       & \textbf{0.0961}          & 0.8081           & \textbf{0.0909}                   & 0.048794 \\
\midrule
\(\beta_8\) & Bias       & 0.0524          & \textbf{-0.0256} & -0.3205  & 0.0033  \\
           & RMSE       & \textbf{0.1215}                   & 0.5017  & 0.3269                   & 0.0587 \\
\bottomrule
\end{tabular}%
}

\label{tab:bias_rmse}
\end{table}
\section{Additional Experimental Results}
\label{sec:Aresults}
\subsection{synthetic datasets:} 
\begin{figure}[ht!]
    \centering
   
    \includegraphics[width=0.754\textwidth]{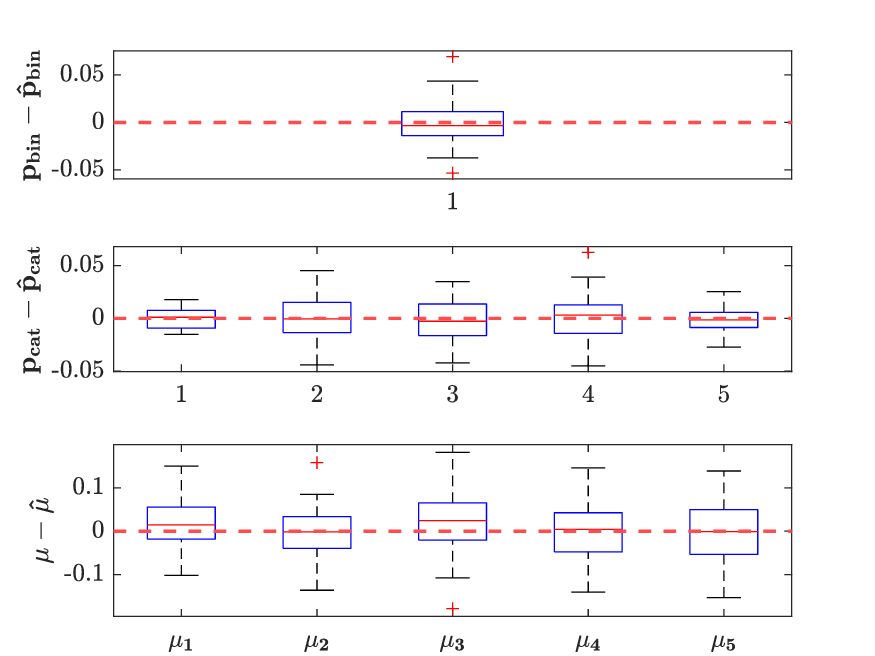} 
    \caption{Boxplot of the difference between the estimate and the ground
truth for SAEM Parameters with 30\% missing data under the MCAR mechanism over  {100 algorithm runs}. 
    }
    \label{fig:bias_analysis}
\end{figure}

\begin{figure}[ht!]
    \centering
    \includegraphics[width=0.4\textwidth]{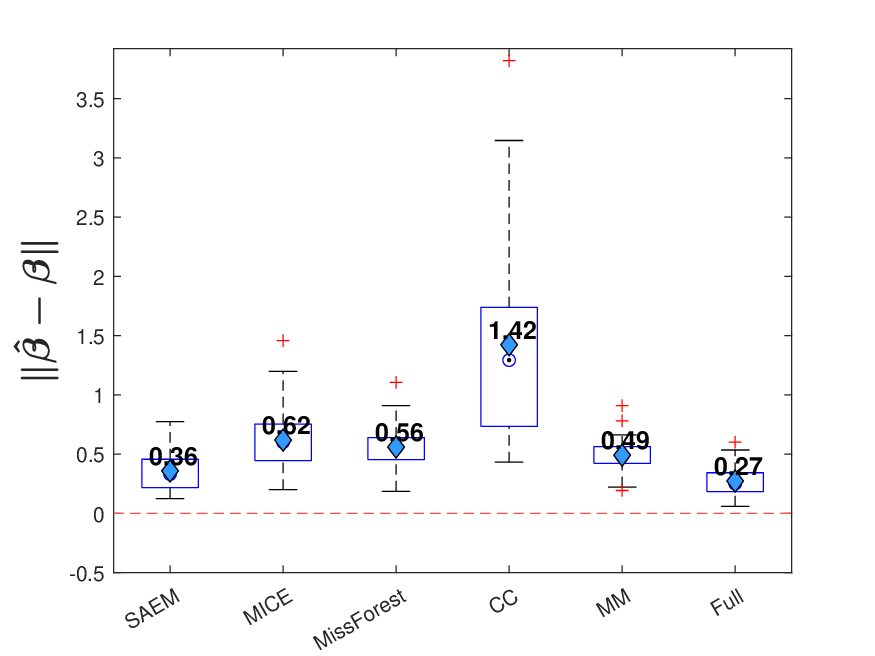} 
    \caption{{Boxplot of the Euclidean norm of the $\bm{\hat{\beta}}$ estimates over  {100 algorithm runs} with 30\% MCAR missing data mechanism.}}
    \label{fig:norm}
\end{figure}

Table \ref{tab:bias_rmse} shows the mean and standard deviation of the bias for the coefficients \(\bm{\beta}\) calculated over  {100 algorithm runs} for 50\% MCAR, along with the corresponding RMSE values. 
 The results indicate that our SAEM algorithm performs well, with low bias and reduced RMSE for most coefficients \( \bm{\beta} \), even in the case of a high ratio of missingness, suggesting high precision. In comparison, \textit{MissForest} shows good performance for certain coefficients \( \bm{\beta} \), but exhibits higher bias for others. On the other hand, \textit{MICE} presents higher bias and RMSE across most coefficients, indicating inferior precision. Fig.~\ref{fig:bias_analysis} presents the difference between the estimated parameters and the ground truth of the SAEM method under a 30\% MCAR mechanism. The estimations for the multinomial probabilities (\(p_{\text{xm}}\)) and the binary probability (\(p_{\text{bin}}\)) for discrete variables, as well as the components of the mean vector \(\ve{\mu}\) for continuous variables, show a significant reduction in bias and variability.
{Fig.~\ref{fig:norm} presents the estimated Euclidean norm of $\bm{\hat\beta}$ under 30\,\% MCAR missingness. This metric provides a global assessment of the estimator’s accuracy across all components of $\bm{\hat\beta}$. The results demonstrate that the SAEM algorithm outperforms the other methods, achieving the smallest Euclidean norm.
 }
\begin{figure}[ht]
    \centering    \includegraphics[width=0.75\textwidth,height=0.55\textwidth,keepaspectratio]{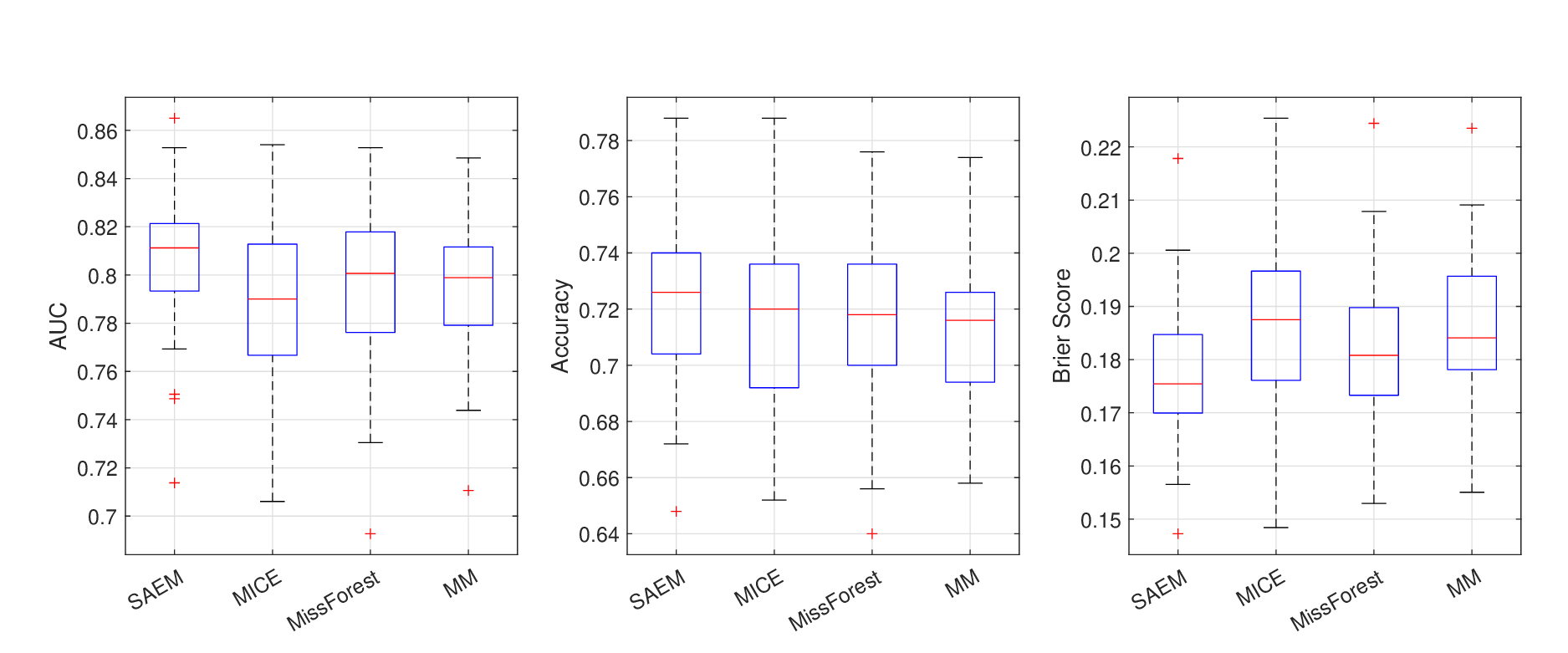} 
    \caption{{Comparison of the Area Under the Curve (AUC), Accuracy, and Brier Score obtained on the test set for each method with 30\% MCAR missing data over  {100 algorithm runs}.}}
    \label{fig:pridect}
\end{figure}
{we focus on the predictive performance. Fig.~\ref{fig:pridect} showcases the superior performance of the \textit{SAEM} algorithm in terms of Area Under the Curve (AUC) and accuracy, as well as a significantly reduced Brier Score. The Brier Score, where lower values indicate better performance, reflects the quality of the probabilistic predictions made by the model. These results, obtained under both MCAR and MAR mechanisms, highlight the effectiveness of our method in handling missing data during the prediction phase. }

\subsection{Real Dataset:}
Tables~\ref{tab:titanic_performance} and~\ref{tab:metrics_comparison_20_40} summarize the predictive performance of the methods presented in the paper on two real-world datasets: the Titanic dataset, which has 12\% naturally occurring missing values, and the Mtcars dataset, into which we injected MCAR missing values at 20\% and 40\% levels. Across both datasets and all missing data scenarios, the SAEM approach consistently outperforms other methods, achieving the highest values for AUC, accuracy, precision, specificity, F1 score, and the lowest Brier score. These results confirm the superior predictive ability of SAEM in realistic settings, whether missing data arise naturally or are introduced artificially. Moreover, the robustness of SAEM to various degrees and mechanisms of missing data underlines its versatility for practical applications. Detailed metric values (mean~$\pm$~standard deviation) are reported in the respective tables.
 {Additionally, Table~\ref{tab:metrics_comparison_50_60} presents the performance of the methods on the Algerian wildfire dataset, with artificially introduced MCAR missing values at 50\% and 60\%. Despite the high missing data rates, the SAEM approach maintains superior predictive performance, achieving the highest values for most evaluation indicators and the lowest Brier scores. This demonstrates the robustness and effectiveness of SAEM in handling datasets with significant missing information, confirming its applicability in real-world scenarios where data incompleteness is a common challenge. Furthermore, Table~\ref{tab:metrics_comparison} presents the results on the same dataset with 30\% missing values introduced via a MAR mechanism. In this scenario, SAEM outperforms all other algorithms across all evaluation metrics, highlighting its exceptional ability to account for missing information patterns that depend on the observed data.}

\begin{table}[h!]
\centering
\scriptsize
\begin{tabular}{lcccc}
\toprule
Metric        & SAEM                  & MICE           & MM             & MissForest     \\ 
\midrule
AUC           & \textbf{0.865 }(0.022) & 0.856 \textbf{(0.021)}  & 0.856 (0.022)  & 0.856 \textbf{(0.021)}  \\
Accuracy      & \textbf{0.813} (0.022)& 0.801 \textbf{(0.021)}  & 0.805 (0.024)  & 0.808 (0.022)  \\
Precision     & \textbf{0.793 (0.041)} & 0.776 (0.042)  & 0.781 (0.058)  & 0.789 (0.045)  \\
Sensitivity   & 0.699 (0.048)         & 0.695 (0.048)  & \textbf{0.704 }(0.051) & 0.694 \textbf{(0.045)}  \\
Specificity   & \textbf{0.889 (0.024)} & 0.870 (0.026)  & 0.870 (0.039)  & 0.883 (0.026)  \\
F1 Score      & \textbf{0.740 (0.037)} & 0.732 (0.036)  & 0.739 (0.036)  & 0.739 (0.036)  \\
Brier Score   & \textbf{0.141} (0.012) & 0.142 (0.011)  & 0.144 (0.011)  & \textbf{0.141 (0.011)}  \\
\bottomrule
\end{tabular}
\caption{{Performance comparison of methods on the Titanic dataset with 12\% naturally missing data (mean ± standard deviation) over  {100 algorithm runs}.}}
\label{tab:titanic_performance}
\end{table}

\begin{table}[h!]
\centering
\scriptsize
\begin{tabular}{lcccc}
Metric & SAEM (20\%) & MICE (20\%) & MM (20\%) & MissForest (20\%) \\ \hline
AUC & \textbf{0.978} (0.077) & 0.817 (0.223) & 0.939 (0.124) & 0.965 \textbf{(0.076)} \\
Accuracy & 0.932 \textbf{(0.083)} & 0.838 (0.194) & 0.898 (0.121) & \textbf{0.936} (0.090) \\
Precision & \textbf{0.946} (0.186) & 0.808 (0.276) & 0.883 (0.208) & 0.932 \textbf{(0.101)} \\
Sensitivity & \textbf{0.909 (0.171)} & 0.809 (0.245) & 0.871 (0.202) & 0.864 (0.186) \\
F1 score & \textbf{0.900 (0.125)} & 0.809 (0.215) & 0.862 (0.153) & 0.899 (0.133) \\
Brier score & \textbf{0.060 (0.077)} & 0.162 (0.194) & 0.100 (0.120) & 0.063 (0.089) \\
\hline
\end{tabular}


\scriptsize

\begin{tabular}{lcccc}
Metric & SAEM (40\%) & MICE (40\%) & MM (40\%) & MissForest (40\%) \\ \hline
AUC & \textbf{0.961} (0.100) & 0.855 (0.193) & 0.943 (0.108) & 0.959 \textbf{(0.057)} \\
Accuracy & \textbf{0.907} (0.122) & 0.836 (0.177) & 0.886 (0.112) & 0.906 \textbf{(0.068)} \\
Precision & \textbf{0.886} (0.193) & 0.796 (0.248) & 0.877 (0.207) & 0.882 \textbf{(0.112)} \\
Sensitivity & \textbf{0.890} (0.194) & 0.832 (0.234) & 0.846 (0.213) & 0.866 \textbf{(0.167)} \\
F1 score & \textbf{0.881} (0.146) & 0.799 (0.191) & 0.855 (0.138) & 0.879 \textbf{(0.109)} \\
Brier score & \textbf{0.071} (0.102) & 0.162 (0.178) & 0.113 (0.112) & 0.078 \textbf{(0.066)} \\
\hline
\end{tabular}

\caption{Performance comparison of methods on the \textit{Mtcars} dataset with 20\% and 40\% MCAR missing data (mean ± standard deviation) across  {100 algorithm runs}.}
\label{tab:metrics_comparison_20_40}
\end{table}
\begin{table}[h!]
\centering
\scriptsize

\begin{tabular}{lcccc}
Metric       & SAEM (50\%)            & MICE (50\%)            & MM (50\%)               & MissForest (50\%)      \\ \hline
AUC          & \textbf{0.966} (\textbf{0.017}) & 0.653 (0.164)         & 0.950 (0.020)           & 0.944 (0.032)          \\
Accuracy     & \textbf{0.913} (\textbf{0.029}) & 0.625 (0.138)         & 0.875 (0.043)           & 0.903 (0.030)          \\
Precision    & \textbf{0.923} (\textbf{0.038}) & 0.669 (0.134)         & 0.888 (0.056)           & 0.913 (0.041) \\
Sensitivity  & \textbf{0.932} (\textbf{0.043}) & 0.642 (0.167)         & 0.890 (0.099)           & 0.923 (0.052)          \\
Specificity  & 0.892 (\textbf{0.044})          & 0.609 (0.160)         & 0.849 (0.096)           & \textbf{0.903} (0.052)  \\
F1 score     & \textbf{0.922} (\textbf{0.027}) & 0.649 (0.140)         & 0.884 (0.049)           & 0.911 (0.027) \\
Brier score  & \textbf{0.068} (\textbf{0.020}) & 0.365 (0.145)         & 0.103 (0.042)           & 0.078 (0.031)          \\ \hline
\end{tabular}

\begin{tabular}{lcccc}
Metric       & SAEM (60\%)            & MICE (60\%)            & MM (60\%)               & MissForest (60\%)      \\ \hline
AUC          & \textbf{0.966} (\textbf{0.020}) & 0.530 (0.109)         & 0.948 (0.026)           & 0.947 (0.039)          \\
Accuracy     & \textbf{0.911} (\textbf{0.030}) & 0.521 (0.087)         & 0.855 (0.042)           & 0.900 (0.032)          \\
Precision    & \textbf{0.905} (\textbf{0.051}) & 0.575 (0.108)         & 0.880 (0.066)           & 0.902 (0.052)          \\
Sensitivity  & \textbf{0.942} (\textbf{0.040}) & 0.512 (0.146)         & 0.864 (0.114)           & 0.922 (0.060)          \\
Specificity  & \textbf{0.877} (\textbf{0.068}) & 0.529 (0.140)         & 0.838 (0.103)           & 0.873 (0.074)          \\
F1 score     & \textbf{0.921} (\textbf{0.026}) & 0.534 (0.115)         & 0.864 (0.052)           & 0.910 (0.033)          \\
Brier score  & \textbf{0.070} (\textbf{0.022}) & 0.473 (0.095)         & 0.115 (0.043)           & 0.085 (0.030)          \\ \hline
\end{tabular}

\caption{{Performance comparison of methods on the Algerian Forest Fires dataset with 50\% and 60\% MCAR missing data (mean $\pm$ standard deviation) over  {100 algorithm runs}.}}
\label{tab:metrics_comparison_50_60}
\end{table}
\begin{table}[h!]
\centering
\scriptsize
\begin{tabular}{lcccc}
\hline
\textbf{Metric} & \textbf{SAEM} & \textbf{MICE} & \textbf{MM} & \textbf{MissForest} \\
\hline
AUC           & \textbf{0.985 (0.015)} & 0.877 (0.135) & 0.970 (0.052) & 0.940 (0.069) \\
Accuracy      & \textbf{0.952 (0.025)} & 0.859 (0.115) & 0.935 (0.043) & 0.911 (0.065) \\
Precision     & \textbf{0.953 (0.034)} & 0.863 (0.098) & 0.937 (0.061) & 0.913 (0.080) \\
Sensitivity   & \textbf{0.964 (0.041)} & 0.905 (0.106) & 0.956 (0.051) & 0.943 (0.065) \\
Specificity   & \textbf{0.941 (0.043)} & 0.794 (0.183) & 0.903 (0.127) & 0.859 (0.169) \\
F1\_score     & \textbf{0.957 (0.023)} & 0.880 (0.088) & 0.944 (0.031) & 0.924 (0.049) \\
Brier\_score  & \textbf{0.046 (0.023)} & 0.139 (0.116) & 0.059 (0.043) & 0.087 (0.063) \\
\hline
\end{tabular}
\caption{{Performance comparison of methods on the Algerian Forest Fires dataset with 30\% MAR missing data (mean $\pm$ standard deviation) over  {100 algorithm runs}.}}
\label{tab:metrics_comparison}
\end{table}

\begin{table}[ht]
\setlength{\tabcolsep}{4pt} 
\centering
\footnotesize 
\begin{tabular}{lcccc}
\toprule
\textbf{Dataset} & \textbf{Samples} & \textbf{Continuous} & \textbf{Binary} & \textbf{Categorical} \\
\midrule
Heart Disease           & 303  & 5  & 2  & 6  \\
Titanic                 & 891  & 2  & 3  & 6  \\
Mtcars         & 32   & 7  & 3  & 1  \\
Algerian Forest Fires   & 244  & 10 & 1  & 2  \\
\bottomrule
\end{tabular}
\caption{{Characteristics of the datasets used, including the number of samples and the distribution of variable types.}}
\label{tab:datasets}
\end{table}

\section{Derivation of the Complete Log-Likelihood}
\label{appendix:loglikelihood}

\begingroup

Assuming that the continuous covariates follow a multivariate Gaussian distribution and that the discrete covariates are modeled by independent multinomial distributions (with Bernoulli distributions as special cases), we define
\[
\bm{\theta}^c = \{ \bm{\mu}, \boldsymbol{\Sigma} \} \quad \text{and} \quad \bm{\theta}^{d}_j = \{ \theta^{d}_j(m) : m = 1, \ldots, M_j \},
\]
where \(\theta^{d}_j(m)\) is the probability of observing level \(m\) for the \(j\)-th discrete variable.

Under the assumptions of i) Independence across observations, ii) independence between the continuous and discrete covariates, and iii) mutual independence among the discrete covariates, 
the joint likelihood for the \(i\)-th observation factorizes as
\begin{equation}
p(y_i, \bm{x}_i; \bm{\theta}) = p(y_i|\bm{x}_i; \boldsymbol{\beta}) \, p(\bm{x}_i^c; \bm{\theta}^c) \, p(\bm{x}_i^d; \bm{\theta}^d).
\label{eq:factorization}
\end{equation}
The first term corresponds to the logistic regression, while the second and third terms are the distributions of the continuous variables (modeled as multivariate Gaussian distribution) and discrete variables, respectively. By
taking the logarithm and summing over all observations yields the complete log-likelihood:
\begin{align}
\mathcal{L}(\bm{\theta}; \bm{y}, \mathcal{X}) 
&= \sum_{i} \log \Bigl( p(y_i, \bm{x}_i; \bm{\theta}) \Bigr) \\
&= \sum_{i} \Bigl\{ \log p(y_i|\bm{x}_i; \boldsymbol{\beta}) 
+ \log p(\bm{x}_i^c; \bm{\theta}^c) 
+ \log p(\bm{x}_i^d; \bm{\theta}^d) \Bigr\}\\
&=\sum_{i} \Biggl\{ y_i \, [1, \bm{x}_i^T]\boldsymbol{\beta} - \log\Bigl(1 + \exp\big([1, \bm{x}_i^T]\boldsymbol{\beta}\big)\Bigr) \\
&+ \frac{1}{2} \Bigl[-h \log(2\pi) - \log |\boldsymbol{\Sigma}| - (\bm{x}_i^c - \bm{\mu})^T \boldsymbol{\Sigma}^{-1} (\bm{x}_i^c - \bm{\mu})\Bigr] \\
&+ \sum_{j=1}^l \log \theta^{d}_j\left( x^j_i \right) \Biggr\}.
\label{eq:complete_likelihood}
\end{align}



\endgroup

\section{Calculation cost}

{Regarding the time complexity of the proposed algorithm, we report the prediction time in Table~.\ref{Testing} and Table~.\ref{tab:exec-timesP_test}. The SAEM algorithm demonstrates remarkable efficiency during the testing phase, achieved through precomputed Gaussian conditionals and Monte Carlo integration over missing covariates. This computational efficiency stems from its ability to leverage pre-optimized statistical approximations while avoiding computationally intensive recalculations. For instance, in
configuration P5 with 4 categorical variables, SAEM achieves testing times of \SI{2.10 \pm 0.20}{s} (Table~\ref{tab:exec-timesP_test}), significantly outperforming MissForest (\SI{38.00 \pm 3.00}{s}) while preserving statistical rigor.}

\begin{table}[ht]
  \centering
  
  \setlength{\tabcolsep}{3pt}
  \renewcommand{\arraystretch}{0.9}
  \scriptsize

  \begin{tabularx}{\columnwidth}{l *{5}{>{\centering\arraybackslash}X}}
    \toprule
    \multirow{2}{*}{\textbf{Estimator}} & \multicolumn{5}{c}{\textbf{Sample Size}} \\
    \cmidrule(lr){2-6}
     & 200 & 500 & 700 & 1000 & 3000 \\
    \midrule
    SAEM       & \SI{0.15 +- 0.06}{} & \SI{0.41 +- 0.11}{} & \SI{0.48 +- 0.07}{} & \SI{0.83 +- 0.14}{} & \SI{3.20 +- 0.38}{} \\
    MICE       & \SI{3.03 +- 2.08}{} & \SI{0.58 +- 1.26}{} & \SI{0.19 +- 0.04}{} & \SI{0.24 +- 0.06}{} & \SI{0.48 +- 0.10}{} \\
    MissForest & \SI{5.74 +- 0.72}{} & \SI{7.63 +- 1.08}{} & \SI{7.02 +- 0.88}{} & \SI{8.90 +- 1.04}{} & \SI{15.58 +- 1.76}{} \\
    \bottomrule
  \end{tabularx}
  \caption{{{Testing Time (mean ± standard deviation in seconds) of Different Algorithms as a Function of Sample Size for 30\% MCAR Missing Data}}}
  \label{Testing}
\end{table}

\begin{table}[ht]
  \centering
  
  \setlength{\tabcolsep}{3pt}       
  
  \renewcommand{\arraystretch}{0.9}
  
  \scriptsize
  \begin{tabularx}{\columnwidth}{l *{5}{>{\centering\arraybackslash}X}}
    \toprule
    \multirow{2}{*}{\textbf{Estimator}} & \multicolumn{5}{c}{\textbf{Configurations}} \\
    \cmidrule(lr){2-6}
     & P1 & P2 & P3 & P4 & P5 \\
    \midrule
    SAEM       & \SI{0.83 \pm 0.14}{} & \SI{0.93 \pm 0.11}{} & \SI{1.21 \pm 0.06}{} & \SI{1.45 \pm 0.15}{} & \SI{2.10 \pm 0.20}{} \\
    MICE       & \SI{0.24 \pm 0.06}{} & \SI{0.27 \pm 0.03}{} & \SI{0.39 \pm 0.06}{} & \SI{0.42 \pm 0.07}{} & \SI{0.48 \pm 0.08}{} \\
    MissForest & \SI{8.90 \pm 1.04}{} & \SI{13.27 \pm 1.82}{} & \SI{23.58 \pm 0.76}{} & \SI{28.50 \pm 2.50}{} & \SI{38.00 \pm 3.00}{} \\
    \bottomrule
  \end{tabularx}
  \caption{{{Testing time (mean ± standard deviation in seconds) for 30\% MCAR Missing Data.
   (P1): 1 binary variable, 1 categorical variable (5 categories), 5 continuous variables.
   (P2): 1 binary variable, 1 categorical variable (5 categories), 10 continuous variables.
   (P3): 1 binary variable, 1 categorical variable (5 categories), 15 continuous variables.
   (P4): 2 binary variables, 2 categorical variables (5 categories each), 5 continuous variables.
   (P5): 4 binary variables, 4 categorical variables (5 categories each), 5 continuous variables.
  }}}
  \label{tab:exec-timesP_test}
\end{table}

\end{document}